\documentstyle[aps,12pt]{revtex}
\begin{document}
\author{E.N. Glass\thanks{%
Permanent address: Physics Department, University of Windsor, Windsor,
Ontario N9B 3P4}}
\address{Physics Department, University of Michigan \\
Ann Arbor, Michigan 48109}
\title{ANGULAR MOMENTUM AND KILLING POTENTIALS }
\date{J. Math. Phys. -- to appear}
\maketitle

\begin{abstract}
When the Penrose-Goldberg (PG) superpotential is used to compute the angular
momentum of an axial symmetry, the Killing potential $Q_{(\varphi )}^{\mu
\nu }$ for that symmetry is needed. Killing potentials used in the PG
superpotential must satisfy Penrose's equation. It is proved for the
Schwarzschild and Kerr solutions that the Penrose equation does not admit a $%
Q_{(\varphi )}^{\mu \nu }$ at finite r and therefore the PG superpotential
can only be used to compute angular momentum asymptotically.\newpage\
\end{abstract}

\section{INTRODUCTION}

In this work computing angular momentum with the use of Killing potentials
is studied for the Schwarzschild and Kerr solutions. Killing potentials are
bivectors $Q^{\mu \nu }$ whose divergence yields a Killing vector. Both
solutions have explicit rotational Killing symmetries, spherical for
Schwarzschild and axial for Kerr, and we have obtained an axial Killing
potential $Q_{(\varphi )}^{\mu \nu }$ for both solutions. We expected to use
that $Q_{(\varphi )}^{\mu \nu }$ in the Penrose-Goldberg (PG) superpotential%
\cite{josh} to compute angular momentum in the same way that $Q_{(t)}^{\mu
\nu }$ has been previously used to compute mass\cite{ed-mark} and found, to
our surprise, that this was not possible.

Killing potentials used in the PG superpotential must satisfy Penrose's
equation\cite{roger}
\begin{equation}
{\bf P}^{\alpha \mu \nu }:=\nabla ^{(\alpha }Q^{\mu )\nu }-\nabla ^{(\alpha
}Q^{\nu )\mu }+g^{\alpha [\mu }Q_{\ \ \ ;\beta }^{\nu ]\beta }=0  \label{pgq}
\end{equation}
such that $\nabla _\beta Q^{\alpha \beta }$ is a Killing vector. Penrose
showed that 10 independent $Q^{\mu \nu }$ exist in Minkowski space, but
there can be no solutions in a general spacetime which has no Killing
symmetries. For Penrose's quasi-local mass integral we exhibit, in the
following section, a Killing potential for the Kerr spacetime which
satisfies (\ref{pgq}) and yields a quasi-local Kerr mass. Unfortunately, one
cannot use the PG superpotential to compute quasi-local angular momentum and
so this work has a {\bf negative} result. It is proved for the Schwarzschild
and Kerr solutions that the Penrose equation does not admit a $Q_{(\varphi
)}^{\mu \nu }$ at finite $r$ and thus the PG superpotential {\bf cannot} be
used to compute angular momentum at finite $r$.

A Newman-Penrose null tetrad for the Kerr solution is given in Appendix A
together with the details of an anti self-dual bivector basis. Bivector
components of the Penrose equation are presented in Appendix B. The
conformal Penrose equation is given in Appendix C. Sign conventions used
here are $2A_{\nu ;[\alpha \beta ]}=A_\mu R_{\ \nu \alpha \beta }^\mu ,$ and
$R_{\mu \nu }=R_{\ \mu \nu \alpha }^\alpha .$

\section{KILLING POTENTIALS}

For Killing vector $k^\alpha $ there is an antisymmetric Killing potential $%
Q^{\alpha \beta }$ such that
\[
k^\alpha =\frac 13\nabla _\beta Q^{\alpha \beta }
\]
It is the Killing potential which is the core of the PG superpotential for
computing conserved Noether quantities such as mass and angular momentum.
The PG superpotential is
\begin{equation}
U^{\alpha \beta }=\sqrt{-g}\frac 12G_{\ \ \mu \nu }^{\alpha \beta }Q^{\mu
\nu },  \label{pgpot}
\end{equation}
where $G_{\ \ \mu \nu }^{\alpha \beta }$ is the negative right and left dual
of the Riemann tensor. In order for
\[
\nabla _\beta U^{\alpha \beta }=\sqrt{-g}G^{\alpha \nu }k_\nu
\]
it is necessary that the Killing potential $Q^{\mu \nu }$ satisfy the
Penrose equation.

The Kerr solution has two Killing vectors, stationary $k_{(t)}^\alpha $ and
axial $k_{(\varphi )}^\alpha $ and the metric, in Boyer-Lindquist
coordinates, is given by
\begin{eqnarray}
g_{\alpha \beta }^{Kerr}dx^\alpha dx^\beta &=&\Psi \ dt^2-(\Sigma /\Delta )\
dr^2+(1-\Psi )2a\sin ^2\theta \ dtd\varphi  \label{kerr} \\
&&\ \ -\Sigma \ d\theta ^2-\sin ^2\theta [\Sigma +(2-\Psi )a^2\sin ^2\theta
]\ d\varphi ^2  \nonumber
\end{eqnarray}
where $R=r-ia\cos \theta $,\quad $\Sigma =R\bar{R}$, \quad $\Delta
=r^2+a^2-2mr$, \quad $\Psi =1-2mr/\Sigma $. The Killing potential for $%
k_{(t)}^\alpha $ is
\begin{equation}
Q_{(t)}^{\alpha \beta }=-\frac 12(RM^{\alpha \beta }+\bar{R}\bar{M}^{\alpha
\beta }).  \label{qt}
\end{equation}
Here $M^{\alpha \beta }$ is an anti self-dual bivector, $M^{*\alpha \beta
}=-iM^{\alpha \beta }$, given in terms of Newman-Penrose null vectors in
Appendix A. One third the divergence of Eq.(\ref{qt}) yields the stationary
Killing vector
\begin{equation}
k_{(t)}^\alpha =n^\alpha +(\Delta /2\Sigma )l^\alpha +(ia\sin \theta /\sqrt{2%
}\Sigma )(\bar{R}m^\alpha -R\bar{m}^\alpha ).  \label{tkvec}
\end{equation}
Direct substitution of $Q_{(t)}^{\alpha \beta }$ in Eq.(\ref{pgq}) verifies
that $Q_{(t)}$ satisfies the Penrose equation. One can now use the
stationary Killing potential with the PG superpotential to compute the mass%
\cite{ed-mark} of the Kerr source:
\begin{equation}
M\,(S^2)=-\frac 1{16\pi }%
%TCIMACRO{\doint }
%BeginExpansion
\displaystyle \oint
%EndExpansion
\limits_{S^2}\sqrt{-g}C_{\ \ \mu \nu }^{\alpha \beta }Q_{(t)}^{\mu \nu }\
dS_{\alpha \beta }  \label{kerrmass}
\end{equation}
where $S^2$ is a closed $t=const,\ r=const$ two surface. The result is $m$
for any $r$ beyond the outer event horizon.

An axial Killing potential for the Kerr solution is given by

\begin{equation}
Q_{(\varphi )}^{\alpha \beta }=Q_1M^{\alpha \beta }+Q_2V^{\alpha \beta }+c.c.
\label{qphi}
\end{equation}
\[
Q_1=\frac{ar\sin ^2\theta }{2\Sigma }(r^2+3a^2\cos ^2\theta ),\ \ Q_2=\frac{%
ir\sin \theta }{\sqrt{2}R}(r^2+3a^2\cos ^2\theta ),
\]
and one third the divergence of $Q_{(\varphi )}^{\alpha \beta }$ yields the
axial Killing vector
\begin{equation}
k_{(\varphi )}^\alpha =-a\sin ^2\theta [n^\alpha +(\frac \Delta {2\Sigma }%
)l^\alpha ]-[\frac{i(r^2+a^2)\sin \theta }{\sqrt{2}\Sigma }](\bar{R}m^\alpha
-R\bar{m}^\alpha ).  \label{phikvec}
\end{equation}
When the Kerr rotation parameter is set to zero, one obtains the
Schwarzschild results
\begin{equation}
Q_{(\varphi )}^{\alpha \beta }=\frac{ir^2\sin \theta }{\sqrt{2}}\ V^{\alpha
\beta }+c.c.  \label{schwqphi}
\end{equation}
\begin{equation}
k_{(\varphi )}^\alpha =-\frac{ir\sin \theta }{\sqrt{2}}\ m^\alpha +c.c
\label{schwphikvec}
\end{equation}
Neither the $Q_{(\varphi )}$ for Kerr nor the $Q_{(\varphi )}$ for
Schwarzschild satisfy the Penrose equation.

\section{NO AXIAL PENROSE SOLUTION}

We will show for the Schwarzschild solution and the Kerr solution that the
Penrose equation does not allow an axial Killing potential at finite $r$.
Penrose's equation\cite{roger}, $\nabla _{A^{\prime }}^{\ \ (A}W_{}^{BC)}=0$
for symmetric spinor $W^{BC}$ (equivalent to the antisymmetric Killing
potential $Q^{\mu \nu }$), was used in linearized theory where Penrose\cite
{penrin2} showed existence of ten independent Killing potentials, one for
each Minkowski Killing vector. In Goldberg's generalization\cite{josh} to a
fully curved metric there is no discussion of the existence of solutions of
the Penrose equation at finite $r$. We know that a solution exists for $%
Q_{(t)}$. It is given in Eq.(\ref{qt}) for the Kerr solution with anti
self-dual components
\begin{equation}
Q_0=0,\ Q_1=-\frac 12R,\ Q_2=0,  \label{kerrtq}
\end{equation}
where
\[
Q^{\mu \nu }=Q_0U^{\mu \nu }+Q_1M^{\mu \nu }+Q_2V^{\mu \nu }+c.c.
\]
We also know that Penrose obtained asymptotic results for angular momentum $%
J.$ For axial symmetry $k_{(\varphi )}$ at the conformal boundary he found $%
J=0$ for Schwarzschild's solution and $J=ma$ for Kerr's, so it is reasonable
to expect a $Q_{(\varphi )}$ for use in the PG superpotential at finite $r$.

The argument presented below assumes that $Q_{(\varphi )}$ exists, goes
through a long set of equations which are the components of the bivector
form of Penrose's equation given in Appendix B, and ends with no possible $%
Q_{(\varphi )}$. To integrate the equations it is assumed that $Q_0,\ Q_1,\ $%
and $Q_2$ are independent of $t$ and $\varphi $, i.e. it is assumed that $%
{\cal L}_\xi Q_{(\varphi )}^{\mu \nu }=0$ where $\xi ^\alpha $ is a Killing
vector that commutes with the Kerr $k_{(t)}^\alpha $ and $k_{(\varphi
)}^\alpha $. If this assumption is false then ${\cal L}_\xi Q_{(\varphi
)}^{\mu \nu }=X^{\mu \nu }$. Penrose's Eq.(\ref{pgq}) with $\nabla _\beta
Q^{\nu \beta }=3k^\nu $ can be written as
\begin{equation}
\nabla _\beta Q^{\mu \nu }=\nabla _{}^{[\mu }Q_{\ \ \beta }^{\nu
]}+3k_{}^{[\mu }\delta _{\ \ \beta }^{\nu ]}.  \label{pgq1}
\end{equation}
Since the Lie and covariant derivatives commute, the non-zero bivector $%
X^{\mu \nu }$ must satisfy
\begin{equation}
\nabla _\beta X^{\mu \nu }=\nabla _{}^{[\mu }X_{\ \ \beta }^{\nu ]}.
\label{nop0}
\end{equation}
The Kerr and Schwarzschild solutions do not admit a non-zero $X^{\mu \nu }$
at finite $r$.

We investigate the existence of $Q_{(\varphi )}$ for the Schwarzschild
solution since the equations are simpler with the Kerr rotation parameter
set to zero but the argument can be extended in a straight forward manner to
the Kerr solution. The null tetrad and spin coeficients given in Appendix A
are used. Penrose's Eq.(\ref{vcomp}) has $n^\alpha $ component
\begin{equation}
L_0=0=\partial _rQ_0,  \label{nop1}
\end{equation}
with solution $Q_0=h(\theta ),$ $h$ an arbitrary function. The $\bar{m}%
^\alpha $ component is
\begin{equation}
M_0=0=\frac 1{\sqrt{2}r}(\partial _\theta Q_0-\cot \theta Q_0),  \label{nop2}
\end{equation}
with solution $Q_0=f(r)\sin \theta $, $f$ arbitrary. The two separate
solutions require
\begin{equation}
Q_0=c_0\sin \theta ,\ \ c_0\ \text{const}.  \label{nop3}
\end{equation}
Equation (\ref{ucomp}) has $l^\alpha $ component
\begin{equation}
N_2=0=r(r-2m)\partial _rQ_2-2mQ_2,  \label{nop4}
\end{equation}
with solution $Q_2=(1-2m/r)h(\theta )$. The $m^\alpha $ component is
\begin{equation}
B_2=0=\frac 1{\sqrt{2}r}(\partial _\theta Q_2-\cot \theta Q_2),  \label{nop5}
\end{equation}
with solution $Q_2=f(r)\sin \theta $. The two solutions for $Q_2$ require
\begin{equation}
Q_2=c_2(1-2m/r)\sin \theta ,\ \ c_2\ \text{const}.  \label{nop6}
\end{equation}
The $n^\alpha $ component of (\ref{ucomp}) is
\begin{eqnarray}
L_2-2B_1 &=&0,  \label{nop7} \\
\partial _rQ_2-\frac 2rQ_2-\frac{\sqrt{2}}r\partial _\theta Q_1 &=&0.
\nonumber
\end{eqnarray}
Using $Q_2$ from (\ref{nop6}) we find
\begin{equation}
Q_1=c_2\sqrt{2}(1-3m/r)\cos \theta +f(r).  \label{nop8}
\end{equation}
We now have functional forms for $Q_0,\ Q_1,\ $and $Q_2.$ The $Q$ components
are further restricted by using the $\bar{m}^\alpha $ component of (\ref
{ucomp}):
\begin{eqnarray}
M_2-2N_1 &=&0,  \label{nop9} \\
\frac 1{\sqrt{2}r}(\partial _\theta Q_2+\cot \theta Q_2)+(1-\frac{2m}r%
)(\partial _rQ_1-\frac 1rQ_1) &=&0.  \nonumber
\end{eqnarray}
Using $Q_2$ from (\ref{nop6}) and $Q_1$ from (\ref{nop8}) we obtain the
equation
\begin{equation}
\frac{c_26\sqrt{2}m}{r^2}\cos \theta +\partial _rf-\frac 1rf=0.
\label{nop10}
\end{equation}
No solution is possible unless one chooses $c_2=0$. Then $Q_1=c_1r$. The $Q$
components are now
\begin{equation}
Q_0=c_0\sin \theta ,\ \ Q_1=c_1r,\ \ Q_2=0.  \label{nop11}
\end{equation}
The $l^\alpha $ component of (\ref{vcomp}) is
\begin{eqnarray}
N_0-2M_1 &=&0,  \label{nop12} \\
\frac 12(1-\frac{2m}r)\partial _rQ_0+\frac m{r^2}Q_0-\frac 1r(1-\frac{2m}r%
)Q_0+\frac{\sqrt{2}}r\partial _\theta Q_1 &=&0.  \nonumber
\end{eqnarray}
Substituting (\ref{nop11}) requires $c_0=0.$ Comparing (\ref{nop11}) and (%
\ref{kerrtq}) one can now see that the only solution possible is the one for
$Q_{(t)}$ given above.

We have proved that, for the Schwarzschild and Kerr solutions, only the
timelike Killing vector $k_{(t)}$ can have a Killing potential that
satisfies the Penrose equation at finite $r$.

\section{NULL INFINITY}

We proceed to solve the Penrose equation at the boundary of Schwarzschild
spacetime. The Schwarzschild solution is given in outgoing null coordinates
as
\begin{equation}
g_{\mu \nu }dx^\mu dx^\nu =(1-2m/r)du^2+2dudr-r^2(d\theta ^2+\sin ^2\theta
d\varphi ^2).  \label{nullschw}
\end{equation}
We use the null tetrad
\begin{eqnarray*}
l_\alpha dx^\alpha &=&du, \\
n_\alpha dx^\alpha &=&\frac 12(1-2m/r)du+dr, \\
m_\alpha dx^\alpha &=&-(r/\sqrt{2})(d\theta +i\sin \theta d\varphi ),
\end{eqnarray*}
and spin coefficients given in Eq.(\ref{spinco}) with Kerr rotation
parameter $a=0$. The general equations for a conformal map are given in
Appendix C. We choose $\Omega =1/r=z.$ On ${%
\hbox{${\cal J}$\kern -.645em {\raise
      .57ex\hbox{$\scriptscriptstyle (\ $}}}}^{+},$ where $z=0,$ the metric
is
\begin{equation}
\hat{g}_{\mu \nu }dx^\mu dx^\nu =-2dudz-(d\theta ^2+\sin ^2\theta d\varphi
^2).  \label{confschw}
\end{equation}
Here the conformal Bondi frame is
\begin{eqnarray*}
\hat{l}_\alpha dx^\alpha &=&du, \\
\hat{n}_\alpha dx^\alpha &=&-dz, \\
\hat{m}_\alpha dx^\alpha &=&-(1/\sqrt{2})(d\theta +i\sin \theta d\varphi ),
\end{eqnarray*}
with non-zero spin coefficients
\[
\hat{\beta}=\frac{\cot \theta }{2\sqrt{2}}=-\hat{\alpha}.
\]
The Penrose equation comprises eight complex equations (\ref{ucomp}), (\ref
{mcomp}), and (\ref{vcomp}) for $\hat{Q}_0,$ $\hat{Q}_1,$ and $\hat{Q}_2$.
Three establish finite values for the $Q$'s on the boundary:
\begin{eqnarray*}
\partial _z\hat{Q}_0 &=&0, \\
\partial _z\hat{Q}_1+\frac 12(\bar{\delta}-2\alpha \hat{)}\hat{Q}_0 &=&0, \\
\partial _z\hat{Q}_2+2(\bar{\delta}\hat{)}\hat{Q}_1 &=&0,
\end{eqnarray*}
where $\hat{D}=-\partial _z,\ \ \hat{\Delta}=\partial _u,\ \ $and on ${%
\hbox{${\cal J}$\kern -.645em {\raise
      .57ex\hbox{$\scriptscriptstyle (\ $}}}}^{+}\ (\hat{\delta}+2s\hat{%
\alpha})\eta =-{\partial \mskip -6.5mu^{\raise 0.46pt\hbox{$%
\scriptscriptstyle\sim$}}\eta }$ for $\eta $ a spin weight $s$ scalar (we
use the original definition\cite{edth} of edth with spin weight opposite to
the helicity of outgoing radiation). In the following a zero superscript
denotes independence of $z$, and ($\hat{Q}_{\ 0}^0,\hat{Q}_{\ 1}^0,\hat{Q}%
_{\ 2}^0$) have spin weights ($1,0,-1$). The remaining five equations on ${%
\hbox{${\cal J}$\kern -.645em {\raise
      .57ex\hbox{$\scriptscriptstyle (\ $}}}}^{+}$ are
\begin{eqnarray}
\partial _u\hat{Q}_{\ 2}^0 &=&0,\ \ \ (a)  \label{atscri} \\
{\bar{\partial}\mskip -6.5mu^{\raise 0.46pt\hbox{$\scriptscriptstyle\sim$}}}%
\hat{Q}_{\ 2}^0 &=&0,\ \ \ (b)  \nonumber \\
{\bar{\partial}\mskip -6.5mu^{\raise 0.46pt\hbox{$\scriptscriptstyle\sim$}}}%
\hat{Q}_{\ 2}^0+2\partial _u\hat{Q}_{\ 1}^0 &=&0,\ \ \ (c)  \nonumber \\
{\partial \mskip -6.5mu^{\raise 0.46pt\hbox{$\scriptscriptstyle\sim$}}}\hat{Q%
}_{\ 0}^0 &=&0,\ \ \ (d)  \nonumber \\
2{\partial \mskip -6.5mu^{\raise 0.46pt\hbox{$\scriptscriptstyle\sim$}}}\hat{%
Q}_{\ 1}^0+\partial _u\hat{Q}_{\ 0}^0 &=&0.\ \ \ (e)  \nonumber
\end{eqnarray}
The solutions are
\begin{eqnarray}
\hat{Q}_{\ 2}^0 &=&k^m\ _{-1}Y_{1m},  \label{qatscri} \\
\hat{Q}_{\ 1}^0 &=&-\frac 12u{\partial \mskip -6.5mu^{\raise 0.46pt%
\hbox{$\scriptscriptstyle\sim$}}}\hat{Q}_{\ 2}^0+f(\theta ,\varphi ),
\nonumber \\
\hat{Q}_{\ 0}^0 &=&\frac 12u^2{\partial \mskip -6.5mu^{\raise 0.46pt%
\hbox{$\scriptscriptstyle\sim$}2}}\hat{Q}_{\ 2}^0-2u{\partial \mskip -6.5mu^{%
\raise 0.46pt\hbox{$\scriptscriptstyle\sim$}}}{\it f}+c^m\ _1Y_{1m},
\nonumber
\end{eqnarray}
where $k^m$ and $c^m$ are complex constants. Here we can go beyond Goldberg%
\cite{josh} and integrate (\ref{atscri}e) since the Schwarzschild null
surfaces are shear-free. The asymptotic Killing vectors are
\begin{eqnarray}
\hat{k}_u &=&\hat{Q}_{\ 1}^0+c.c.  \label{katscri} \\
\Omega \hat{k}_\theta &=&c.c.(\hat{Q}_{\ 2}^0)  \nonumber \\
\Omega \hat{k}_\varphi &=&\hat{Q}_{\ 2}^0  \nonumber
\end{eqnarray}
The supertranslations of the BMS group have a full function's worth of
freedom in $\hat{Q}_{\ 1}^0$ but at the Schwarzschild boundary $f(\theta
,\varphi )$ is restricted to four parameters for ordinary translations and ${%
\partial \mskip -6.5mu^{\raise 0.46pt\hbox{$\scriptscriptstyle\sim$}}}f=0$.

The solution of the Penrose equation for $Q_{(t)}$ is contained above. The
non-zero anti self-dual component of Eq.(\ref{qt}) is $Q_1=-r/2$ or $\hat{Q}%
_1=-1/2.$ This solution coincides with the values $k^m=0,\ c^m=0,$ and $%
f(\theta ,\varphi )=-1/2$.

Now lets take the asymptotic solutions found above in (\ref{qatscri}) and (%
\ref{katscri}) and use them to construct a Killing potential $Q_{(\varphi )}$%
. Thus our candidate has the form
\begin{equation}
Q_{(\varphi )}^{\mu \nu }=(r^2Q_{\ 2}^0)V^{\mu \nu }+c.c.  \label{qcand}
\end{equation}
For the Schwarzschild solution we compute the divergence:
\begin{equation}
\frac 13\nabla _\nu Q_{(\varphi )}^{\mu \nu }=-[rQ_{\ 2}^0]m^\mu +[\frac r{%
\sqrt{2}}(\partial _\theta +\cot \theta )Q_{\ 2}^0]l^\mu +c.c  \label{qcand1}
\end{equation}
Equating with
\[
k_{(\varphi )}^\mu =-\frac{ir\sin \theta }{\sqrt{2}}\ m^\mu +c.c
\]
yields
\[
Q_{\ 2}^0=\frac{i\sin \theta }{\sqrt{2}}.
\]
The $l^\mu $ term in (\ref{qcand1}) vanishes when the complex conjugate is
added. We have constructed the Killing potential which was already given
above as Eq.(\ref{schwqphi}). The anti self-dual components are
\begin{equation}
Q_0=0,\ \ Q_1=0,\ \ Q_2=(i/\sqrt{2})r^2\sin \theta .  \label{schwqphicomp}
\end{equation}
Of the twelve terms entering the Penrose equation (defined in Appendix B)
four are non-zero for the components of Eq.(\ref{schwqphicomp}):\\

$L_2=i\sqrt{2}r\sin \theta ,\ \ \ \ N_2=(i/\sqrt{2})(3m-r)\sin \theta ,$

$M_2=ir\cos \theta ,\ \ \ \ \ \ B_1=(i/\sqrt{2})r\sin \theta .$\\\\Although $%
Q_2$ has the $r^2$ dependence that one expects for an asymptotic solution
and the angular dependence dictated by $k_{(\varphi )}$, the components of
Eq.(\ref{ucomp}), particularly $N_2=0$, show directly that this Killing
potential fails to satisfy the Penrose equation.

\section{CONCLUSION}

To find a Killing potential one can write the divergence equation relating $%
k^\alpha $ and $Q^{\alpha \beta }$ as a 3-form relation, one third the
exterior derivative of dual $Q$ equal to the dual of $k_\alpha dx^\alpha $%
\[
\frac 13d\ ^{*}Q\ =\ ^{*}(k_\alpha dx^\alpha )
\]
and then integrate (if possible). We have seen that not just any Killing
potential can be used in the PG superpotential but only one which satisfies
Penrose's equation. Although a $Q_{(\varphi )}^{\alpha \beta }$ whose
divergence yielded the axial Killing vector was presented for the Kerr
solution, it could not be used to compute quasi-local angular momentum
although asymptotically it yields $ma$. It has been shown that a $%
Q_{(\varphi )}^{\alpha \beta }$ {\bf cannot} be found for either the Kerr or
Schwarzschild solutions which will satisfy the Penrose equation in curved
space and so the PG superpotential cannot be used to compute quasi-local
angular momentum.

Some interesting questions remain. What are the complete integrability
conditions for the Penrose equation? What is the physical reason that no
quasi-local Killing potential for rotational symmetry can satisfy the
Penrose equation?

\begin{center}
{\bf ACKNOWLEDGMENTS}
\end{center}

We thank Bob Geroch for questioning the existence of Killing potentials
which satisfy the Penrose equation and David Garfinkle for many stimulating
discussions. This work was partially supported by an NSERC of Canada grant.

\appendix

\section{NULL TETRAD AND BIVECTORS}

A Newman-Penrose tetrad $(l^\alpha ,n^\alpha ,m^\alpha ,\bar{m}^\alpha )$
for the Kerr metric (\ref{kerr}) with $l^\alpha $ and $n^\alpha $ as
principal null vectors is chosen as
\begin{eqnarray}
l^\alpha \partial _\alpha &=&\frac 1\Delta [(r^2+a^2)\partial _t+\Delta
\partial _r+a\partial _\varphi ],  \label{tetrad} \\
n^\alpha \partial _\alpha &=&\frac 1{2\Sigma }[(r^2+a^2)\partial _t-\Delta
\partial _r+a\partial _\varphi ],  \nonumber \\
m^\alpha \partial _\alpha &=&\frac 1{\sqrt{2}\bar{R}}[ia\sin \theta \partial
_t+\partial _\theta +\frac i{\sin \theta }\partial _\varphi ],  \nonumber
\end{eqnarray}
where $R=r-ia\cos \theta $,$\ \ \Sigma =R\bar{R}$, $\ \Delta =r^2+a^2-2mr$.
The non-zero spin coefficients and Weyl tensor component are
\begin{eqnarray}
\rho &=&-1/R,\qquad \mu =-\Delta /(2\Sigma R),\qquad \tau =-ia\sin \theta /(%
\sqrt{2}\Sigma ),  \label{spinco} \\
\pi &=&ia\sin \theta /(\sqrt{2}R^2),\qquad \gamma =\mu +(r-\text{m}%
)/(2\Sigma ),  \nonumber \\
\beta &=&\cot \theta /(2\sqrt{2}\bar{R}),\qquad \alpha =\pi -\bar{\beta}%
,\qquad \psi _2=-m/R^3.  \nonumber
\end{eqnarray}
A basis of anti self-dual bivectors is given by
\begin{equation}
U^{\mu \nu }=2\bar{m}^{[\mu }n^{\nu ]},\qquad M^{\mu \nu }=2l^{[\mu }n^{\nu
]}-2m^{[\mu }\bar{m}^{\nu ]},\qquad V^{\mu \nu }=2l^{[\mu }m^{\nu ]}.
\label{bivecs}
\end{equation}
Their inner products are $U^{\mu \nu }V_{\mu \nu }=\bar{U}^{\mu \nu }\bar{V}%
_{\mu \nu }=2,\ \ M^{\mu \nu }M_{\mu \nu }=\bar{M}^{\mu \nu }\bar{M}_{\mu
\nu }=-4,$ and all others zero. As a basis, they satisfy the completeness
relation
\begin{equation}
\frac 12(g^{\alpha \beta \mu \nu }+i\eta ^{\alpha \beta \mu \nu })=U^{\alpha
\beta }V^{\mu \nu }+V^{\alpha \beta }U^{\mu \nu }-\frac 12M^{\alpha \beta
}M^{\mu \nu },  \label{comp}
\end{equation}
where $g^{\alpha \beta \mu \nu }=g^{\alpha \mu }g^{\beta \nu }-g^{\alpha \nu
}g^{\beta \mu }$, and $\frac 12\eta ^{\alpha \beta \mu \nu }$ is the dual
tensor. It is useful to list their covariant derivatives:
\begin{eqnarray}
\nabla _\beta U^{\mu \nu } &=&-2U^{\mu \nu }a_\beta +M^{\mu \nu }b_\beta
,\qquad a_\beta =\epsilon n_\beta +\gamma l_\beta -\alpha m_\beta -\beta
\bar{m}_\beta ,  \label{bideriv} \\
\nabla _\beta M^{\mu \nu } &=&-2U^{\mu \nu }c_\beta +2V^{\mu \nu }b_\beta
,\qquad b_\beta =\pi n_\beta +\nu l_\beta -\lambda m_\beta -\mu \bar{m}%
_\beta ,  \nonumber \\
\nabla _\beta V^{\mu \nu } &=&\ \ 2V^{\mu \nu }a_\beta -M^{\mu \nu }c_\beta
,\ \ \ \ \ \ c_\beta =\kappa n_\beta +\tau l_\beta -\rho m_\beta -\sigma
\bar{m}_\beta .  \nonumber
\end{eqnarray}
\newpage\

\section{THE PENROSE EQUATION}

Equation (\ref{pgq}), which a Killing potential must satisfy in order to be
valid for use in the PG superpotential, can be written in terms of anti
self-dual bivectors with the definition
\begin{equation}
Q^{\mu \nu }=Q_0U^{\mu \nu }+Q_1M^{\mu \nu }+Q_2V^{\mu \nu }+c.c.
\label{bivq}
\end{equation}
Substituting the bivector expansion into (\ref{pgq}) provides equations for
the components $Q_0,Q_1,Q_2$ which can be most simply written with the use
of twelve terms:\\\\$L_0=(D-2\epsilon )Q_0-2\kappa Q_1,\ \ L_1=DQ_1-\kappa
Q_2+\pi Q_0,\ L_2=(D+2\epsilon )Q_2+2\pi Q_1,$\\\\$N_0=(\Delta -2\gamma
)Q_0-2\tau Q_1,\ N_1=\Delta Q_1-\tau Q_2+\nu Q_0,\ N_2=(\Delta +2\gamma
)Q_2+2\nu Q_1,$\\\\$M_0=(\delta -2\beta )Q_0-2\sigma Q_1,\ M_1=\delta
Q_1-\sigma Q_2+\mu Q_0,\ M_2=(\delta +2\beta )Q_2+2\mu Q_1,$\\\\$B_0=(\bar{%
\delta}-2\alpha )Q_0-2\rho Q_1,\ \ B_1=\bar{\delta}Q_1-\rho Q_2+\lambda
Q_0,\ \ B_2=(\bar{\delta}+2\alpha )Q_2+2\lambda Q_1.$\\\\Here $D=l^\alpha
\nabla _\alpha $, $\Delta =n^\alpha \nabla _\alpha $, $\delta =m^\alpha
\nabla _\alpha $. The Penrose equation has the following $U_{\mu \nu }$, $%
M_{\mu \nu }$, and $V_{\mu \nu }$ components respectively:
\begin{equation}
l^\alpha (3N_2)+n^\alpha (L_2-2B_1)-m^\alpha (3B_2)-\bar{m}^\alpha
(M_2-2N_1)=0.  \label{ucomp}
\end{equation}
\begin{equation}
l^\alpha (M_2-2N_1)+n^\alpha (B_0-2L_1)+m^\alpha (2B_1-L_2)+\bar{m}^\alpha
(2M_1-N_0)=0.  \label{mcomp}
\end{equation}
\begin{equation}
l^\alpha (N_0-2M_1)+n^\alpha (3L_0)-m^\alpha (B_0-2L_1)-\bar{m}^\alpha
(3M_0)=0.  \label{vcomp}
\end{equation}
If $Q^{\mu \nu }$ is to be a Killing potential for $k^\mu $ then its
divergence must satisfy
\begin{equation}
3k^\mu =l^\mu (N_1+M_2)-n^\mu (B_0+L_1)-m^\mu (B_1+L_2)+\bar{m}^\mu
(N_0+M_1)+c.c.  \label{divq}
\end{equation}
\newpage\

\section{THE CONFORMAL PENROSE EQUATION}

For asymptotically simple spacetimes with future null infinity ${%
\hbox{${\cal J}$\kern -.645em {\raise
      .57ex\hbox{$\scriptscriptstyle (\ $}}}}^{+}$ we follow Penrose and
Rindler\cite{penrin1} case (iv) to conformally map from the physical metric $%
g_{\alpha \beta }$ to the unphysical metric $\hat{g}_{\alpha \beta }:$%
\begin{equation}
\hat{g}_{\alpha \beta }=\Omega ^2g_{\alpha \beta }  \label{gmap}
\end{equation}
with the spinor basis mapping as $\hat{o}_A=o_A$, \ $\hat{\iota}_A=\Omega
\iota _A$. $\Omega =0$ defines the future null boundary with $\nabla _\alpha
\Omega $ a null vector tangent to the generators of ${%
\hbox{${\cal J}$\kern -.645em {\raise
      .57ex\hbox{$\scriptscriptstyle (\ $}}}}^{+}.\ $It follows from the map
of the spinor basis that the tetrad derivatives transform as
\begin{equation}
\hat{D}=\Omega ^{-2}D,\ \ \hat{\delta}=\Omega ^{-1}\delta ,\ \ \hat{\Delta}%
=\Delta .  \label{derivmap}
\end{equation}
The spin coefficients conformally map as\\

$\hat{\kappa}=\Omega ^{-3}\kappa ,\qquad \qquad \hat{\rho}=\Omega ^{-2}\rho
-\Omega ^{-3}D\Omega ,$

$\hat{\sigma}=\Omega ^{-2}\sigma ,\qquad \qquad \hat{\tau}=\Omega ^{-1}\tau
-\Omega ^{-2}\delta \Omega ,$

$\hat{\epsilon}=\Omega ^{-2}\epsilon ,\qquad \qquad \hat{\alpha}=\Omega
^{-1}\alpha -\Omega ^{-2}\bar{\delta}\Omega ,$

$\hat{\beta}=\Omega ^{-1}\beta ,\ \ \qquad \ \hat{\gamma}=\gamma -\Omega
^{-1}\Delta \Omega ,$

$\hat{\nu}=\Omega \nu ,\qquad \qquad \ \hat{\mu}=\mu +\Omega ^{-1}\Delta
\Omega ,$

$\hat{\lambda}=\lambda ,\qquad \ \qquad \ \hat{\pi}=\Omega ^{-1}\pi +\Omega
^{-2}\bar{\delta}\Omega .$\\ \\ Since the Killing potential obeys the
conformal transformation $\hat{Q}^{\alpha \beta }=\Omega ^{-1}Q^{\alpha
\beta }$, it's anti self-dual bivector components map as
\begin{equation}
\hat{Q}_0=Q_0,\ \ \hat{Q}_1=\Omega Q_1,\ \ \hat{Q}_2=\Omega ^2Q_2.
\label{qmap}
\end{equation}
The twelve terms in Appendix B which comprise the components of the Penrose
equation map as\\ \\ $\hat{L}_0=\Omega ^{-2}L_0,\ \ \hat{L}_1=\Omega
^{-1}L_1+Q_0(\Omega ^{-2}\bar{\delta}\Omega )+Q_1(\Omega ^{-2}D\Omega ),\\ %
\\ \hat{L}_2=L_2+2Q_1(\Omega ^{-1}\bar{\delta}\Omega )+2Q_2(\Omega
^{-1}D\Omega ),\ \ \hat{N}_0=N_0+2Q_0(\Omega ^{-1}\Delta \Omega
)+2Q_1(\Omega ^{-1}\delta \Omega ),\\ \\ \hat{N}_1=\Omega N_1+Q_1(\Delta
\Omega )+Q_2(\delta \Omega ),\ \ \hat{N}_2=\Omega ^2N_2,$\\ \\ $\hat{M}%
_0=\Omega ^{-1}N_0,\ \ \hat{M}_1=M_1+Q_0(\Omega ^{-1}\Delta \Omega
)+Q_1(\Omega ^{-1}\delta \Omega ),\\ \\ \hat{M}_2=\Omega M_2+2Q_1(\Delta
\Omega )+2Q_2(\delta \Omega ),\ \ \hat{B}_0=\Omega ^{-1}B_0+2Q_0(\Omega ^{-2}%
\bar{\delta}\Omega )+2Q_1(\Omega ^{-2}D\Omega ),\\ \\ \hat{B}%
_1=B_1+Q_1(\Omega ^{-1}\bar{\delta}\Omega )+Q_2(\Omega ^{-1}D\Omega ),\ \
\hat{B}_2=\Omega B_2.$\\ \\ Finally, by direct substitution of the twelve
terms above into equations (\ref{ucomp}), (\ref{mcomp}), and (\ref{vcomp})
we find the anti self-dual components of the Penrose equation conformally
transform as
\begin{equation}
(\hat{B}2)=(B2),\ \ (\hat{B}3)=\Omega ^{-1}(B3),\ \ (\hat{B}4)=\Omega
^{-2}(B4).  \label{penmap}
\end{equation}
This result is confirmed by the conformal maps
\begin{equation}
{\bf P}^{\alpha \mu \nu }=\Omega ^{-3}{\bf P}^{\alpha \mu \nu }  \label{pmap}
\end{equation}
and
\begin{equation}
\hat{U}_{\mu \nu }=\Omega ^3U_{\mu \nu },\ \ \hat{M}_{\mu \nu }=\Omega
^2M_{\mu \nu },\ \ \hat{V}_{\mu \nu }=\Omega V_{\mu \nu }.  \label{bivecmap}
\end{equation}

\newpage\

\end{document}